\documentclass[twocolumn,showpacs,amsmath,amssymb,amsfonts]{revtex4}
\usepackage{epsfig} 
\usepackage{dcolumn}% Align table columns on decimal point
\usepackage{bm}% bold math
\usepackage{float}

\newcommand{\be}{\begin{equation}} 
\newcommand{\en}{\end{equation}}
\newcommand{\bea}{\begin{eqnarray}}
\newcommand{\ena}{\end{eqnarray}}

\newcommand{\hbo}{\hbox to 1 true cm {\hfill } } 
\newcommand{\tr}{\hbox{tr}}
\newcommand{\tre}{\hbox{Re tr}}

\begin{document}

\title{ Improved actions and asymptotic scaling \\ 
        in lattice Yang-Mills theory }

% Force line breaks with \\

\author{Kurt Langfeld}
 \email{kurt.langfeld@plymouth.ac.uk} 

\affiliation{%
School of Mathematics \& Statistics, University of Plymouth \\
Plymouth, PL4 8AA, UK }%

\date{(revised) September 5,2007}% It is always \today, today,

\begin{abstract}
Improved actions in SU(2) and SU(3) lattice gauge theories are 
investigated with an emphasis on asymptotic scaling. 
A new scheme for tadpole improvement is proposed. The standard  but 
heuristic tadpole improvement emerges from a mean field approximation 
from the new approach. Scaling is investigated by means of the 
large distance static quark potential. Both, the generic and the new 
tadpole scheme yield significant improvements on asymptotic scaling 
when compared with loop improved actions. A study of the 
rotational symmetry breaking terms, however, reveals that only the 
new improvement scheme efficiently eliminates the leading irrelevant 
term from the action.
\end{abstract}

%Use showkeys class option if keyword
\maketitle

\section{ Introduction:  }

At the beginning of lattice gauge theories, Wilson pointed out that 
it is important to maintain exact gauge invariance for 
finite lattice spacings $a$ thus enforcing gauge invariance 
in the critical limit of the lattice model. The minimal choice for 
an action which satisfies this criterion is nowadays known as the 
Wilson action~\cite{Wilson:1974sk}. Relying on the concept of 
universality, any lattice action which incorporates the
correct symmetries in the continuum limit should work in principle. 
However, some lattice actions do give better approximations for 
continuum results for coarser lattices. This issue is central for 
computer simulations and has led to a continuous development of the so-called 
improved and perfect actions~\cite{Symanzik:1983dc,Symanzik:1983gh,
Luscher:1985zq,Weisz:1982zw,Weisz:1983bn,Iwasaki:1985we,Iwasaki:1996sn,
Lepage:1995ph,GarciaPerez:1996hi,Hasenfratz:1993sp,Hasenfratz:1997ft,
DeGrand:1995ji,DeGrand:1995jk,vanBaal:1996cz,deForcrand:1999bi}. 
The basic idea is to add terms to the action which are 
{\it irrelevant } in the continuum limit, but which give 
better approximations at finite lattice 
spacing~\cite{Symanzik:1983dc,Symanzik:1983gh}. Different 
proposals have been made for such actions on the basis of perturbation 
theory~\cite{Luscher:1985zq,Weisz:1982zw,Weisz:1983bn} or using 
renormalisation group techniques~\cite{Hasenfratz:1993sp,Hasenfratz:1997ft, 
DeGrand:1995ji,DeGrand:1995jk,deForcrand:1999bi}. 
It is widely accepted that the so-called tadpole improvement is important 
for good properties of these actions~\cite{Lepage:1995ph,GarciaPerez:1996hi}. 
To our knowledge, however, a systematic study of different implementations 
of tadpole improvement has not yet been carried out. 

\vskip 0.3cm
In the context of computer simulations, an extrapolation of data to 
the limit of vanishing lattice spacing is necessary. Such extrapolations 
can be made very trustworthy if a relation to an ab initio continuum 
calculation can be established. If we use the string tension 
$\sigma $ as the fundamental energy scale of Yang-Mills theory, 
the perturbative treatment of continuum $SU(N_c)$ Yang-Mills theory predicts 
the dependence of the lattice spacing on $\beta = 2N_c/g^2$ (with 
$g$  the bare gauge coupling) to be: 
\be
\ln \left[ \sigma a^2 \left(\beta\right) \right] \, = \, - \,
\frac{4 \pi^2}{\beta_0} \, \beta \, + \,
\frac{2 \beta_1}{\beta_0^2} \, \ln  \beta  \, + \, 
c_\sigma \, + \, {\cal O}(1/\beta ) 
\label{eq:tension}
\en
There, the 1-loop and 2-loop coefficients 
\be 
\beta _0 \; = \; \frac{11\, N_c^2 }{6} \; , \hbo 
\beta _1 \; = \; \frac{17\, N_c^4 }{12} 
\label{eq:i1}
\en
are universal. The dimensionless parameter $c_\sigma $ depends on the 
observable and must be determined by non-perturbative methods such as lattice 
simulations: any physical mass scale (call it $m$) in units of the 
reference scale is independent of the lattice spacing for 
sufficiently large $\beta $ and is obtained from  
\bea 
\ln \left( \frac{m^2}{ \sigma } \right) &=&
\ln \left[ m^2 a^2 \left(\beta\right) \right] \; - \; 
\ln \left[ \sigma a^2 \left(\beta\right) \right] 
\label{eq:i2} \\
&=& c_m \; - \; c_\sigma  \; + \; {\cal O}(1/\beta ) \; . 
\nonumber
\ena 
Let $N$ denote the number of lattice points in one direction of the 
lattice. In actual lattice simulations, $\beta $ cannot be chosen 
too large if we wish to work with reasonable lattice sizes,  
$N \, a(\beta )$. Using the Wilson lattice action, it turns out 
that these values of $\beta $ are still too small to  observe 
perturbative scaling: for moderate $\beta $ values, large corrections to 
the scaling function (\ref{eq:tension}) are observed. 
Note, however, that the ratio $m^2 a^2 /\sigma  a^2$ is 
almost independent of the lattice spacing at these $\beta $ values 
which let us reliably calculate low energy observables. 
This property, called 'scaling' in the literature, must not be 
confused with ``asymptotic scaling'', i.e.~perturbative scaling, 
which is the focal point of the present paper. 

\vskip 0.3cm 
In this paper, a new tadpole improved action is proposed. The 
construction of this action offers a new understanding of the 
otherwise heuristic ``derivation'' of the standard tadpole action 
commonly used in simulations nowadays. We will find that 
tadpole improvement is highly important for the approach to asymptotic 
scaling for reasonably sized lattices. Finally, a thorough study of 
rotational symmetry breaking effects obtained from the static quark 
potential will reveal that the standard tadpole action does not 
cancel the leading order irrelevant terms of the action. 
The numerical data suggest that a complete cancellation might be 
achieved by means of our new action.

\section{Action and Improvement }

\subsection{ Standard  tadpole improvement }

Because of the non-linear relation between the link field 
$U_\mu (x)$ and the continuum gauge potential, lattice perturbation 
theory suffers from large tadpole contributions which, however, 
must cancel for an extrapolation to the continuum limit. 
That these tadpole contributions are indeed large can be easily 
anticipated from the expectation value of the link in Landau gauge, 
$U^\Omega _\mu (x)$. 
A naive expansion with respect to the lattice spacing, i.e.,  
\bea 
\Bigl\langle \frac{1}{N_c} \tr U^\Omega _\mu (x) \Bigr\rangle &=&
\Bigl\langle 1 \; - \; \frac{1}{2 N_c} \tr \, 
A^2_\mu (x) \,a^2  + \ldots \Bigr\rangle 
\label{eq:ai1} \\
&=&
1 \; + \; {\cal O} (a^2) \; ,  
\nonumber 
\ena 
implies that this expectation value should be of order $1$. 
Actual simulations show, however, that the latter expectation value 
is at most of order $0.8$ although $\sigma a^2 $ is as small as $0.05$. 
In order to improve the approach to the continuum limit, one 
considers the ratio between the link and its expectation value. 
One assumes that the deviation from unity now provides a better definition 
of the (continuum) gauge field $A_\mu (x)$: 
\be 
U^\Omega _\mu (x) \; / \;  \frac{1}{N_c} \langle \tr U^\Omega _\mu (x) 
\rangle  \; = \;  1 \; + \; i \, A_\mu (x) \,a  \; - \; \ldots \; .
\label{eq:ai2}
\en
Since Landau gauge fixing is problematic because of the Gribov 
problem, an ad hoc description for tadpole improvement has 
become standard: defining 
\be 
\widetilde{U}_\mu (x) \; = \; U_\mu (x) / u_0 \; , 
\label{eq:ai3}
\en
where $u_0$ is the fourth root of the plaquette expectation value, 
each link field in a lattice operator should be replaced by 
$\widetilde{U}_\mu (x) $. Note that this procedure is heuristic, and 
that many other choices for tadpole improvement exist: one could 
also choose for $u_0$ the 8th root of the expectation value of 
the $2 \times 2$ planar Wilson loop. Nevertheless, the prescription 
outlined above has become standard.

\subsection{ Motivation of the new action } 

Let us consider a quadratic (planar) Wilson loop of side length $s$ 
with an orientation specified by $\mu $, $\nu $ located at site $x$. 
A naive expansion of this operator 
yields (see e.g.~(2) of~\cite{GarciaPerez:1996hi}): 
\begin{eqnarray}
\frac{1}{N_c} \; \tre \; W_{\mu \nu }(x)  &=& 
1 \; - \; \frac{1}{N_c} \; \tre \Bigl[ \, 
\frac{1}{2} \, O_4 \, s^4 
\label{eq:ai4} \\
&-& \frac{1}{24} \, O_6 \, s^6 \; + \; \ldots \Bigr] 
\nonumber  \\ 
O_4(x) &=& F_{\mu \nu } F_{\mu \nu }(x) \; , 
\label{eq:ai5} \\ 
O_6 (x) &=&  (D_\mu F_{\mu \nu })^2(x) \; + \; 
(D_\nu F_{\mu \nu })^2 (x) \; . 
\nonumber  
\end{eqnarray}
The subscripts $\mu$, $\nu$ at $O_{4,6}$ have been suppressed. 
Already terms of order ${\cal O}(s^6)$ break 
of rotational symmetry. Choosing the minimal length $s=a$, $W$ coincides with 
the (minimal) plaquette, which is the only term in the Wilson action.  

\vskip 0.3cm 
As outlined in the previous subsection, numerically 
the expansion (\ref{eq:ai4}) converges badly for reasonable 
lattice sizes. Without any simulation, this fact can be also understood 
from continuum perturbation theory: although 
manifestly gauge invariant, in lattice regularisation the high energy 
modes are cutoff at a 
momentum scale $\Lambda _{UV} \approx \pi /a $. It is well known that 
in cutoff regularisations expectation values such as $O_4$ and $O_6$ 
in (\ref{eq:ai5}) diverge with the cutoff: 
$$ 
\langle O_4 \rangle \; \propto \; \Lambda _{UV}^4 \; \propto 
\; a^{-4} \; , \hbo 
\langle O_6 \rangle \; \propto \; \Lambda _{UV}^6 \; \propto \; a^{-6} \; . 
$$
The origin of these divergences are quantum fluctuation of the order of 
the cutoff scale. Obviously, these fluctuations invalidate 
the expansion (\ref{eq:ai4}) (choose $s=a$ for the moment). 
However, they do not spoil the calculation of physical observables 
well below the cutoff scale (as will be detailed below). 
Note, however, that if the desired goal is to match with 
{\it asymptotic scaling} provided by continuum Yang-Mills theory, 
an expansion such as (\ref{eq:ai4}) should cover high energy modes too. 

\vskip 0.3cm 
One choice for such an action is obtained by replacing {\it all } 
operators $O_n$ of the action by 
\be 
\bar{O}_n \; = \; O_n \; - \; \langle O_n \rangle \; . 
\label{eq:rel}
\en 
Only the deviation of the gauge invariant operator from its 
(potentially) divergent expectation value contributes to the action. 
In practice, this construction can be realized by considering the 
ratio between the Wilson loop and its expectation value. 
Using (\ref{eq:ai4}), one can show to all orders that in this case the 
operators $O_n$ only appear in the combination (\ref{eq:rel}). 

\vskip 0.3cm 
Without resorting to the naive expansion anymore, we now  assume 
that the above ratio has a sensible expansion with respect to $s$: 
\bea 
\tre \; W_{\mu \nu }(x)  &/& \langle \tre \; W_{\mu \nu }(x) \rangle 
\; = \;  1 \; - \; \frac{1}{N_c} \; \tre \Bigl[ \, 
\nonumber \\
&&\frac{1}{2} \, \bar{O}_4 \, s^4 \; - \; 
\frac{1}{24} \, \bar{O}_6 \, s^6 \; + \; \ldots \Bigr] \; . 
\label{eq:ai6} 
\ena 
Note that the term $ \bar{O}_4 $ gives rise to the continuum action 
proportional to $F^2$. The subject of improvement is to remove 
terms of higher order in $s$ from the action. 

\vskip 0.3cm 
A popular choice (Symanzik improvement) is to use a rectangular 
$1\times 2$ loop. This scheme invokes yet another expansion, i.e., 
of the rectangular loop with respect to $s$ similar to the one in 
(\ref{eq:ai6}), and relies on a matching of the expansion 
coefficients to eliminate the irrelevant terms. 
Here, we are going to use a $2\times 2$ quadratic loop which 
is a scale transform of the plaquette. The motivation 
for this choice is that we need to invoke only one type of expansion 
evaluated at two values for the scale parameter. 
The hope is that, because of the scale relation between the 
$1\times 1$ and $2\times 2$ loops, the cancellation of irrelevant 
terms is more complete at finite values of the lattice 
spacing where the naive Taylor expansion (such as (\ref{eq:ai6})) 
appears unjustified. 

\vskip 0.3cm 
Assuming that the expansion (\ref{eq:ai6}) is valid at least for 
$s \le 2a$, we use the expansion for $s=a$ (plaquette) and 
$s=2a$ ($2\times 2$ Wilson loop) to get rid of the irrelevant terms. 
Defining 
\bea 
\bar{P}_{\mu \nu }(x) &=&
\tre \; W^{1\times 1}_{\mu \nu }(x)  \; /  \; \langle \tre \; 
W^{1\times 1}_{\mu \nu }(x) \rangle \; , 
\label{eq:ai7} \\
\bar{P}^{(2)}_{\mu \nu }(x) &=&
\tre \; W^{2\times 2}_{\mu \nu }(x)  \; /  \; \langle \tre \; 
W^{2\times 2}_{\mu \nu }(x) \rangle \; . 
\nonumber 
\ena
we choose for the action 
\be 
S \; = \; \beta \; \sum _{\mu > \nu, x} \Bigl[ \kappa _1 \; 
\bar{P}_{\mu \nu }(x) \; + \;  \kappa _2 \; \bar{P}^{(2)}_{\mu \nu }(x) 
\Bigr] \; . 
\label{eq:ai8} 
\en
Using the expansion (\ref{eq:ai6}) for $s=a$ and $s=2a$, we are led to 
\bea 
\kappa _1 \; + \; 16 \; \kappa _2 &=&  1 \; , 
\label{eq:ai9} \\
\kappa _1 \; + \; 64 \; \kappa _2 &=& 0 \; . 
\label{eq:ai10} 
\ena 
This first line ensures compatibility with continuum Yang-Mills theory 
whereas the choice of the second line eliminates the order $a^6$ terms. 
The solution of the latter set of equations is given by 
\be 
\kappa _1 \; = \; \frac{4}{3} \; , \hbo 
\kappa _2 \; = \; - \; \frac{1}{48} \; . 
\label{eq:ai11} 
\en 
The present improvement scheme eliminates from the action contributions 
from tadpole loops. The main purpose for this elimination is that  
these loops are absent in the ab initio continuum 
formulation of Yang-Mills theory. The impact of these loops is therefore 
to spoil proper scaling which is familiar from continuum perturbation 
theory. I point out that, once the tadpole contribution were eliminated, 
further improvements might be achieved by invoking the standard perturbative 
improvement scheme. We leave 
such an investigation to future work. Here, we will 
justify by numerical calculations that the new action (without further 
perturbative improvements) already gives rise to much better scaling 
properties.

\subsection{ Comparison with the standard tadpole improved action}

Let us assume that we are dealing with an action which features 
the plaquette and the $2 \times 2 $ Wilson loop. In the case 
of standard tadpole improvement, the rule (\ref{eq:ai3}) implies 
that 
\bea
\bar{P}_{\mu \nu }(x) &=&
 \frac{1}{N_c} \, \tre \; W^{1\times 1}_{\mu \nu }(x)  \; /  \; u_0^4 
\nonumber \\ 
&=&
\tre \; W^{1\times 1}_{\mu \nu }(x)  \; /  \; \langle \tre \; 
W^{1\times 1}_{\mu \nu }(x) \rangle \; , 
\nonumber \\ 
\bar{P}^{(2)}_{\mu \nu }(x) &=&
 \frac{1}{N_c} \, \tre \; W^{2\times 2}_{\mu \nu }(x)  \; /  \;  u_0^8  . 
\nonumber 
\ena 
While for our new action the numerical burden is a self-consistent 
calculation of the expectations values 
$$ 
\langle \tre \; W^{1\times 1}_{\mu \nu }(x) \rangle \; , \hbo 
\langle \tre \; W^{2\times 2}_{\mu \nu }(x) \rangle \; ,
$$
standard tadpole improvement appears as an approximation to this 
numerical problem: There, only 
$$ 
\langle \tre \; W^{1\times 1}_{\mu \nu }(x) \rangle 
$$
is calculated self-consistently, and the expectation value of the 
$2 \times 2 $ Wilson loop is obtained with the help of the mean field 
approximation (in Landau gauge):
\bea 
\left\langle \frac{1}{N_c} \, \tre \; W^{2\times 2}_{\mu \nu }(x) 
\right\rangle 
& \approx & \; u_0^8 \; = \; (u_0^4)^2 \; 
\nonumber \\ 
& \approx & 
\left\langle \frac{1}{N_c} \, 
\tre \; W^{1\times 1}_{\mu \nu }(x) \right\rangle ^2 \; . 
\nonumber 
\ena 
Having identified the standard approach as an approximation to the 
present scheme, the crucial question is whether the properties of our  
action fully justify the higher level of numerical 
sophistication. The remaining two sections will answer this question.

\section{Numerical simulation setup}

\subsection{ Thermalisation \label{sec:therm} }

\begin{figure*}[t]
\centerline{  
\epsfxsize=7cm 
\epsffile{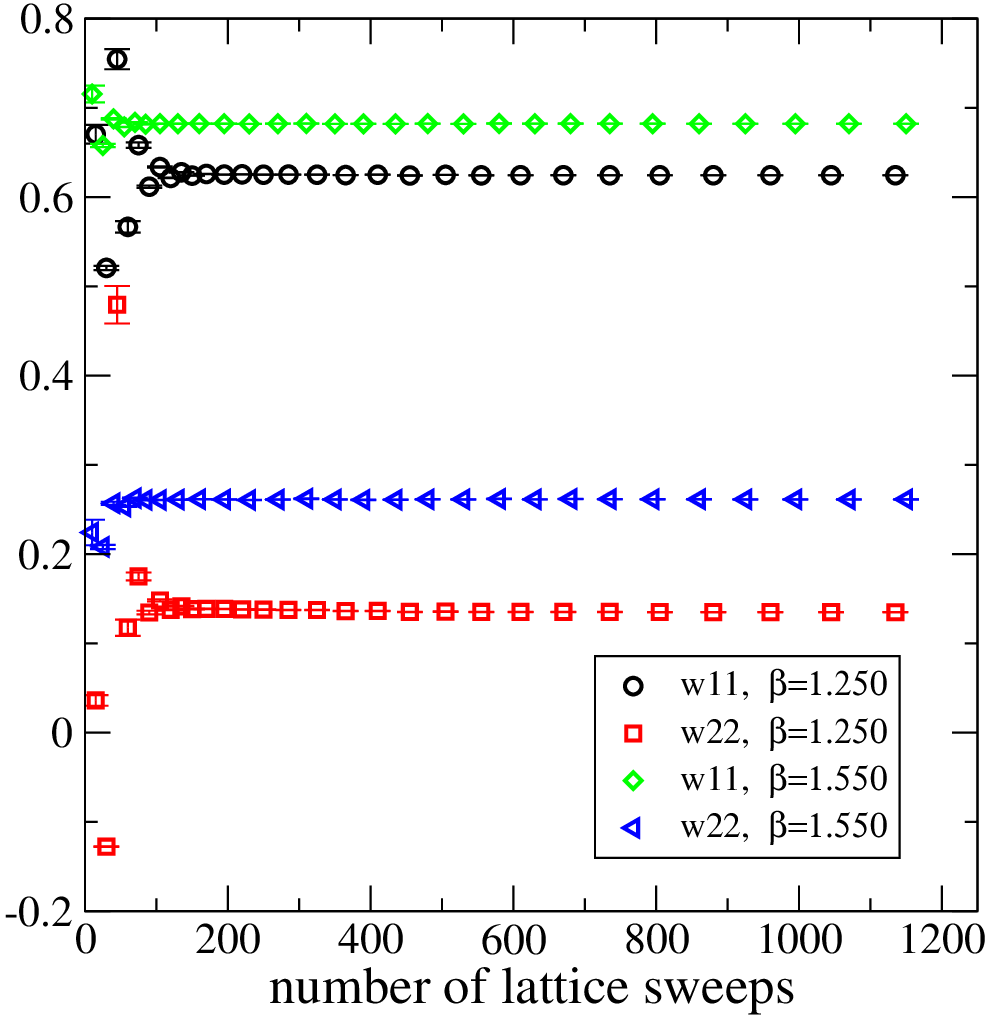} \hspace{.5cm}
\epsfxsize=7cm 
\epsffile{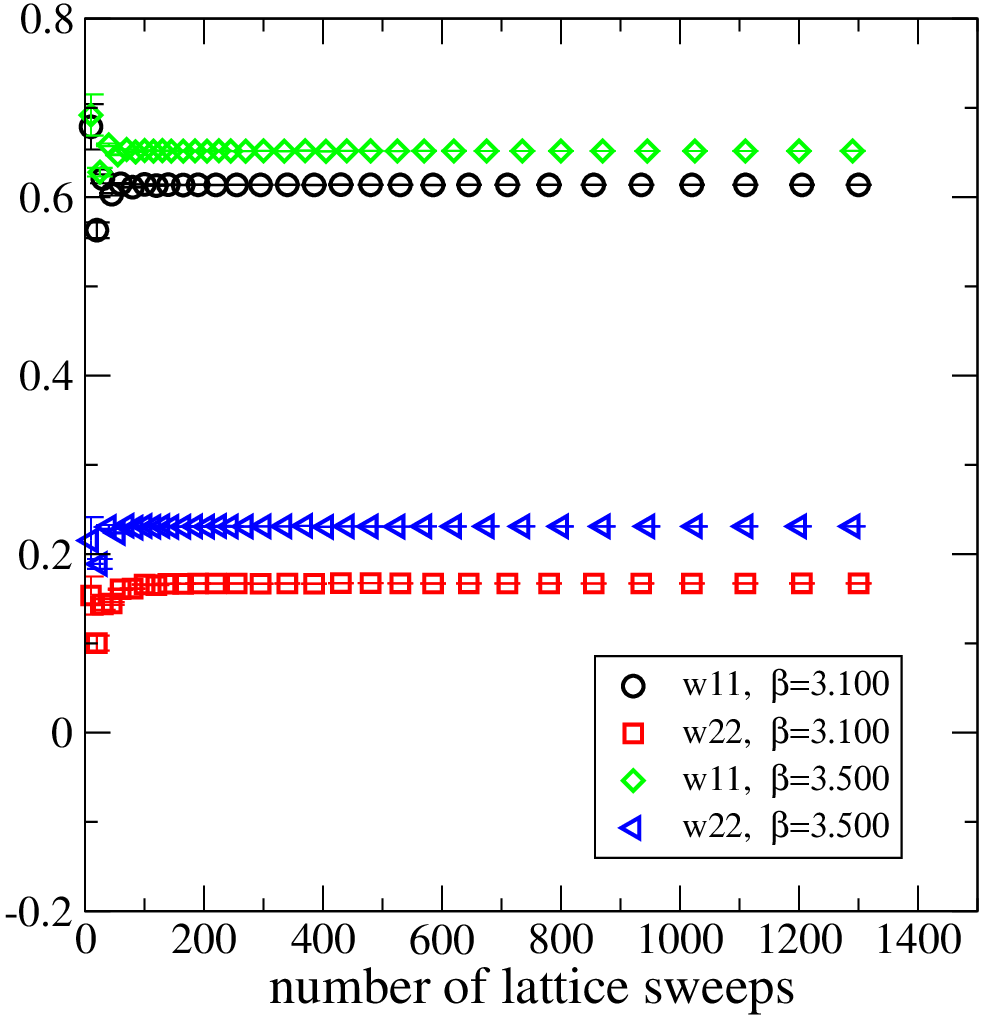} 
} 
\vskip 0.3cm
\caption{ \label{fig:t1} Convergence of the action parameters 
$w_{11}$ and $w_{22}$ as a function of the number of Monte-Carlo sweeps: 
SU(2) (left) and SU(3) (right). 
}
\end{figure*}
The dynamical degrees of freedom are the $SU(N_c)$ matrices $U_\mu (x)$ 
which are associated with the links of a $N^4$ cubic lattice. The 
partition function is given by 
\bea 
Z &=& \int {\cal D} U_\mu \; \exp \{ S[U](w_{11},w_{22}) \} 
\label{eq:t1} \\ 
S[U](w_{11},w_{22}) &=& \beta \; \sum _{\mu < \nu, x} 
\Bigl[ \frac{4}{3 \, w_{11} (\beta )} \; \tre \, W_{\mu \nu }^{1 \times 1} (x) 
\nonumber \\ 
&-& \frac{1}{48 \, w_{22} (\beta )} \; \tre \, W_{\mu \nu }^{2 \times 2} 
(x) \Bigr] \; , 
\label{eq:t2} 
\ena
where the action $S$ depends on the expectation values: 
\bea  
w_{11} (\beta ) &=& \langle  \tre \, W_{\mu \nu }^{1 \times 1} (x) \rangle 
\; , 
\label{eq:t3} \\
w_{22} (\beta ) &=& \langle  \tre \, W_{\mu \nu }^{2 \times 2} (x) 
\rangle \; . 
\nonumber 
\ena 
Expanding the expectation values of the latter equations 
in terms of their functional integrals, we arrive 
at a set of two non-linear equations which determine the two unknown 
parameters $w_{11} (\beta ) $ and $w_{22} (\beta )$: 
\bea 
w_{11} (\beta ) &=& \frac{1}{Z} \int {\cal D} U_\mu \; 
\tre \, W_{\mu \nu }^{1 \times 1} (x) \; 
\nonumber \\ 
&& \hbo \exp \{ S[U](w_{11},w_{22}) \} \; , 
\label{eq:t4} \\ 
w_{22} (\beta )&=& \frac{1}{Z} \int {\cal D} U_\mu \; 
\tre \, W_{\mu \nu }^{2 \times 2} (x)  
\nonumber \\ 
&& \hbo \exp \{ S[U](w_{11},w_{22}) \} \; . 
\label{eq:t5} 
\ena 
Before we can start to accumulate statistically independent lattice 
configurations $\{U_\mu \}$ for each $\beta $, we  must solve the 
latter set of equations for $w_{11} (\beta ) $ and $w_{22} (\beta )$,  
and we must generate a ``statistically important'' configuration 
by means of thermalisation.

\vskip 0.3cm 
While the reader is invited to develop their own methodology for this task, 
we here briefly outline our approach which serves the purpose. 
It appears to be quite natural to solve the set of equations 
(\ref{eq:t4},\ref{eq:t5}) and to generate the thermalized configuration 
within one process. We here used a simple iterative procedure: denoting 
$w^{(n)}_{ii} (\beta )$, $i=1,2$, by the approximate solutions to 
$w_{ii} (\beta )$ of the $n$th iteration, better approximations 
are generated by 
\bea 
w^{(n+1)}_{11} (\beta ) &=& \frac{1}{Z} \int {\cal D} U_\mu \; 
\tre \, W_{\mu \nu }^{1 \times 1} (x) 
\nonumber \\ 
&& \hbo \exp \{ S[U](w^{(n)}_{11},w^{(n)}_{22}) \} \; , 
\label{eq:t6} \\ 
w^{(n+1)}_{22} (\beta ) &=& \frac{1}{Z} \int {\cal D} U_\mu \; 
\tre \, W_{\mu \nu }^{2 \times 2} (x) \; 
\nonumber \\ 
&& \hbo \exp \{ S[U](w^{(n)}_{11},w^{(n)}_{22}) \} \; . 
\label{eq:t7} 
\ena 
As starting points for the iteration we chose the naive tree level values 
\be 
w^{(0)}_{11} (\beta ) \; = \; 1 \; , \hbo 
w^{(0)}_{22} (\beta ) \; = \; 1 \; . 
\label{eq:t8} 
\en
In order to monitor the convergence of the above iteration, we introduce 
the error 
\be 
\epsilon  ^{(n+1)} \; = \; \vert \, w^{(n+1)}_{22} (\beta ) \; - \; 
w^{(n)}_{22} (\beta ) \, \vert  \; . 
\label{eq:t9} 
\en
It turns out that measuring $w^{(n)}_{22} (\beta )$ is sufficient 
for monitoring convergence. In practice, the integrals 
in (\ref{eq:t6},\ref{eq:t7}) are not calculated exactly. Only 
Monte-Carlo estimates $\widetilde{w}^{(n)}_{ii} (\beta )$ 
with statistical errors $\sigma ^{(n)}_{ii} (\beta ) $ are available. 
At the beginning of the iteration, it does not make sense to 
obtain  a high precision estimate for an anyhow un-converged 
value of $\widetilde{w}^{(n)}_{ii} (\beta )$. We therefore adopted 
the following procedure: at the start of the iteration, only $10$ 
iterations are used to obtain the estimates $\widetilde{w}^{(n)}_{ii} 
(\beta)$ and their statistical errors $\sigma ^{(n)}_{ii} (\beta ) $. 
As soon as the error of convergence reaches the order of the 
statistical error, i.e. 
\be 
\epsilon  ^{(n)} \; \approx \; \sigma ^{(n)}_{22} (\beta ) \; , 
\label{eq:t10}
\en 
the number of iterations which are used for the estimators is 
increased by $10$. The iteration stops when 
$\epsilon  ^{(n)} $ (and therefore also $ \sigma ^{(n)}_{22} (\beta )$) 
drops below a certain number which specifies the precision to be 
achieved for the parameters. 
Figure~\ref{fig:t1} shows the ``thermalisation history'' of the parameters 
$\widetilde{w} ^{(n)}_{11} (\beta )/N_c$ and 
$ \widetilde{w} ^{(n)}_{22} (\beta )/N_c$ 
as a function of the total number lattice sweeps performed. Data are shown 
for $\beta = 1.25$ (SU(2)) and $\beta = 3.10$ (SU(3)), which will turn out 
to correspond to a rather coarse lattice, and for $\beta = 1.55$ (SU(2)) 
and $\beta =3.5$ (SU(3)), which are in 
the scaling regime. After an initial oscillation, the estimators 
rapidly converge. Note that the spacing between two data points in 
figure~\ref{fig:t1} shows the number of lattice sweeps which were 
needed to estimate the integrals (\ref{eq:t6},\ref{eq:t7}). 

\vskip 0.3cm
\begin{figure*}[t]
\centerline{  
\epsfxsize=7cm 
\epsffile{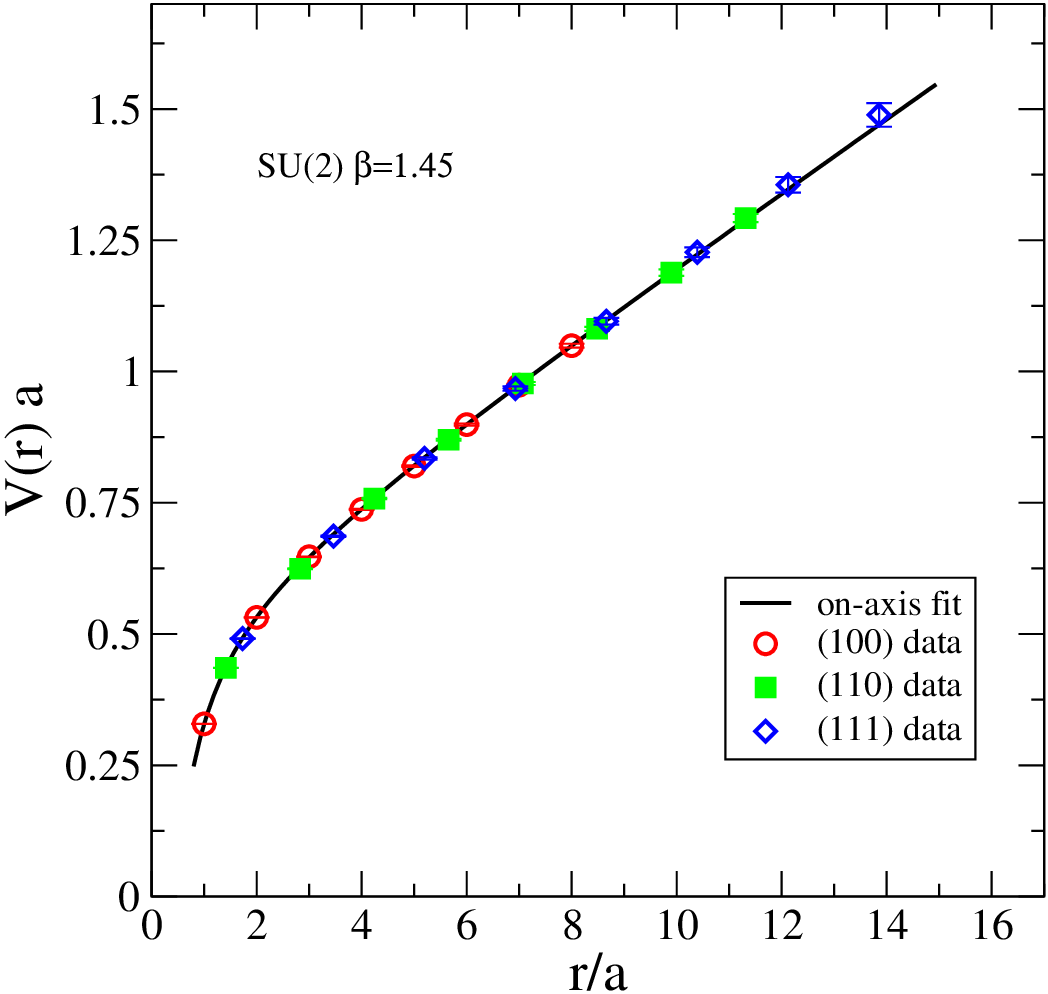} \hspace{.5cm}
\epsfxsize=7cm 
\epsffile{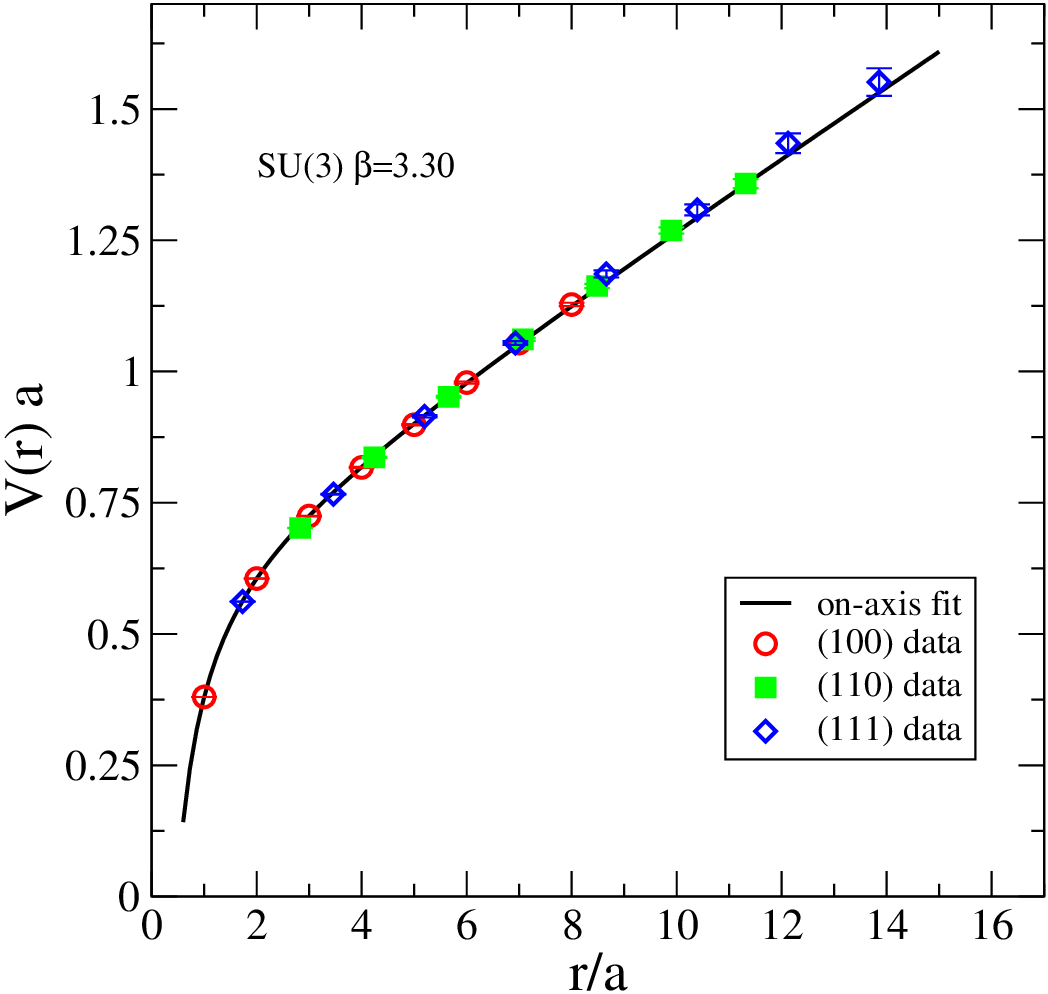} 
} 
\vskip 0.3cm
\caption{ \label{fig:p1} Static quark anti-quark potential using the 
present improved action for SU(2) (left) and SU(3) (right). 
}
\end{figure*}
In particular for small values of $\beta $, several solutions of 
the non-linear equations (\ref{eq:t4},\ref{eq:t5}) might exist. 
If one chooses to perform simulations in this regime of parameter space

\subsection{Static quark potential } 

To investigate scaling, we will express the lattice spacing 
in units of the string tension $\sigma $ in order to calculate $a(\beta )$. 
The static quark potential $V(r)$ is so obtained from planar 
Wilson loops. These loops are of size $r \times t$, and the spatial 
links have been smeared to enhance the overlap with the 
quark anti-quark ground state. For the smearing procedure, we consider 
the spatial hypercube for a given time $t$: spatial links belonging to 
this cube are then cooled with respect to the 3-dimensional Wilson action. 
Cooling is performed by visiting each link of the lattice and 
replacing it the (normalised) sum of the adjacent staples. 
Instead of the Wilson action, one could use the 3-dimensional 
version of the present action rather than Wilson's action. It turns out, 
however, that this choice is more time-consuming and does not produce 
better overlaps. 
10 cooling sweeps through the lattice are performed to obtain one set 
of smeared links. Time like links are unaffected by smearing. 
The advantages of this smearing procedure are that it is easy to implement, 
it is fast compared with other smearing techniques and it is known to 
yield excellent overlap for SU(2) and  SU(3) gauge 
theories~\cite{Langfeld:2003ev} and even for more exotic gauge theories 
such as for G(2)~\cite{Greensite:2006sm}. 

\vskip 0.3cm 
In practice, the Wilson loops are fitted to a straight line: 
\be 
- \; \ln \; \Bigl\langle \tre \, W^{r \times t} _{\mu \nu } 
\Bigr\rangle \; = \; V(r) \; t \; + \; \hbox{const.} \; ,  
\label{eq:p1} 
\en 
where only data with $t \ge t_\mathrm{low}$ are included. This suppresses 
the contribution from excited states. Because of the overlap 
enhancement, choosing $ t_\mathrm{low} = 2a $ is sufficient: the 
linear $t$-fit represents the data with a $\chi^2/\hbox{dof} \approx 2$ 
or better for the $\beta $ ranges explored in the present paper. 
We checked that larger values $ t_\mathrm{low} $ yield the same 
potential. 

\vskip 0.3cm 
In order to explore rotational symmetry breaking effects by the 
underlying lattice, ``off-axis'' distances for the quark anti-quark pair 
are considered as well. Potentials corresponding to cristalographic 
directions 
$$ 
(100) \hbo \hbox{(on-axis)}\;,  \hbo  \hbo (110) \hbo (111) 
$$ 
are taken into account. Our final result for the static potential 
is shown in figure~\ref{fig:p1}. A fit of the on-axis data to 
\be 
V(r) \; = \; V_0 \; - \; \frac{\alpha }{r} \; + \; \sigma \; r 
\label{eq:p2} 
\en 
is shown as well. The results of the fit for SU(2) and SU(3) are summarised 
in the table below: 

\bigskip 
\begin{center} 
\begin{tabular}{c|cccccc}
   & $ \beta $ &  $N_\mathrm{conf}$ & $V_0 \, a$  & $\alpha $ & 
   $\sigma \, a^2$ &  $\chi^2/\hbox{dof}$ \\ \hline 
SU(2) & $1.45$ & $400$ & $0.527(2) $ & $0.267(1)$ & $0.0695(6)$ & $1.1$ \\
SU(3) & $3.30$ & $800$ & $ 0.631(1)$ & $0.317(1)$ & $0.0666(3)$ & $1.2$ \\
\end{tabular} 
\end{center} 

\bigskip 
Here, $N_\mathrm{conf}$ denotes the number of independent lattice 
configurations used to estimate the Wilson loop expectation values. 
Priority has been given to the SU(3) simulations because of their 
relevance for QCD. Because we are using an improved action with 
very good rotational symmetry (see section~\ref{sec:rot}), the point 
$r=a$ can be included in the potential fit without hampering 
the value for $\chi^2/\hbox{dof}$. 

A lattice sweep consists out of a Cabbibo-Marinari update followed by 
$4$ reflections (for SU(2)) and $5$ (for SU(3)), respectively. Each reflection 
replaces the actual configuration by another one which possess the same 
action. We observe that this process enhances the ergodicity of the 
algorithm: auto-correlations are reduced and a speed-up of 
thermalisation is observed. $20$ ``dummy'' lattice sweeps are performed 
until the configuration is used for measurements. 
Especially for small $\beta $ values, a smaller number of dummy sweeps 
might be sufficient. There is room for a further fine-tuning of the 
algorithm. 
Note that these sweeps are carried out for fixed $\kappa _{1,2}$ 
in (\ref{eq:ai8}) the values of which were determined during the 
initial stage of thermalisation.

\section{Asymptotic scaling with improved actions  }

\subsection{ The Wilson action - a case study }

\begin{figure}[t]
\begin{center}
\epsfxsize=8cm 
\epsffile{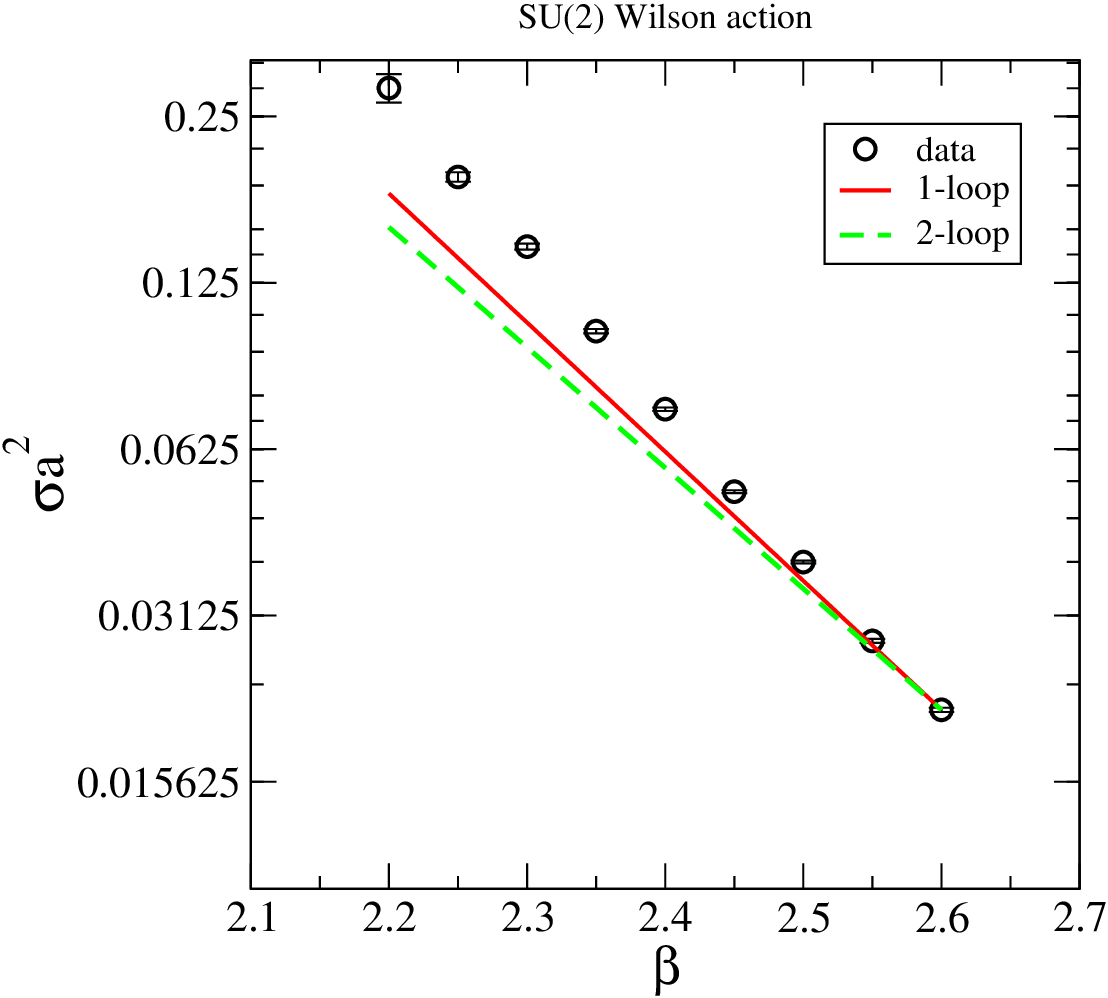} %\hspace{.5cm}

\epsfxsize=8cm 
\epsffile{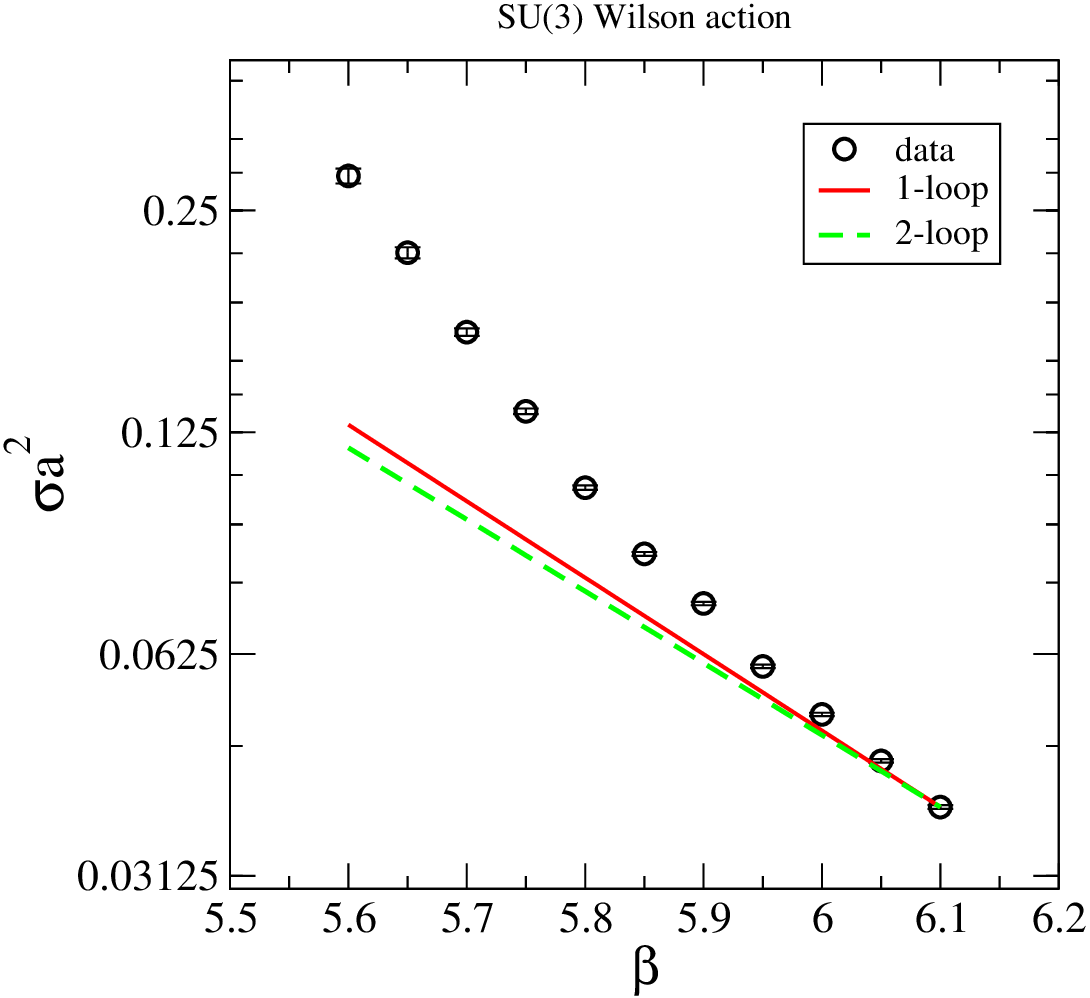} 
\end{center}
\vskip 0.3cm
\caption{ \label{fig:as1} Scaling function $\sigma a^2(\beta)$ 
for a $16^4$ lattice using Wilson action for SU(2) (top) 
and for SU(3) (bottom) gauge theories. Asymptotic scaling 
according to (\ref{eq:as2},\ref{eq:as3}) is shown as well. 
}
\end{figure}
\begin{table}[t]
\begin{center}
\begin{tabular}{|l|c|l|l|c|l|} \hline 
SU(2) &  $ \beta $ &  $\sigma a^2 $ & SU(3) &  $ \beta $ &  $\sigma a^2 $ \\
\hline 
      &  $2.20$ &  $ 0.28(1) $  &   & $  5.60 $   & $ 0.278(6) $  \\
      &  $2.25$ &  $ 0.194(4)$  &   & $  5.65 $   & $ 0.219(4) $  \\
      &  $2.30$ &  $ 0.145(2)$  &   & $  5.70 $   & $ 0.171(2) $  \\
      &  $2.35$ &  $ 0.1022(9)$ &   & $  5.75 $   & $ 0.133(1) $  \\
      &  $2.40$ &  $ 0.0738(5)$ &   & $  5.80 $   & $ 0.1051(7) $ \\
      &  $2.45$ &  $ 0.0523(3)$ &   & $  5.85 $   & $ 0.0854(5) $ \\
      &  $2.50$ &  $ 0.0390(2)$ &   & $  5.90 $   & $ 0.0731(4) $ \\
      &  $2.55$ &  $ 0.0281(2)$ &   & $  5.95 $   & $ 0.0601(3) $ \\
      &  $2.60$ &  $ 0.0211(2)$ &   & $  6.00 $   & $ 0.0517(2) $ \\
      &         &               &   & $  6.05 $   & $ 0.0447(2) $ \\
      &         &               &   & $  6.10 $   & $ 0.0387(2) $ \\ \hline
\end{tabular} 
\end{center}
\caption{ \label{tab:as1} Measured scaling function $\sigma a^2(\beta)$ 
for a $16^4$ lattice using the Wilson action for SU(2) and SU(3) 
gauge theories. }
\end{table}
The Wilson action has been widely studied and is widely used even nowadays. 
It has been known, however, for a long time 
that large deviations of the lattice spacing $a(\beta )$ from the 
perturbative scaling (\ref{eq:tension}) are common with this type 
of action. The purpose of the present subsection is to quantify 
this statement. 

\vskip 0.3cm 
The partition function employing Wilson action is given by 
\bea 
Z &=& \int {\cal D} U_\mu \; \exp \{ S_\mathrm{wil}[U] \} 
\label{eq:as1} \\ 
S_\mathrm{wil}[U] &=& \beta \; \sum _{\mu < \nu, x} 
\frac{1}{N_c} \; \tre \, W_{\mu \nu }^{1 \times 1} (x) \; . 
\label{eq:tas2} 
\ena
Using the techniques outlined in the previous section, 
we calculated the static quark potential and the scaling function 
$\sigma a^2 (\beta )$. The results, obtained from $N_\mathrm{conf}=800$ 
independent configurations on a $16^4$ lattice, are summarised in 
table~\ref{tab:as1}. Finite size effects are expected at the 
1\% level if the side length of the lattice exceeds 
$1.5 \,$fm~\cite{Bali:1992ab}. For a $16^4$ lattice, 
finite size effects therefore play a minor role as long as 
$\sigma \, a^2 > 0.04 $. 

\vskip 0.3cm 
Figure~\ref{fig:as1} visualises the data of table~\ref{tab:as1}. 
In order to bring out any onset of asymptotic scaling, these data 
are compared with the perturbative scaling function 
at 1-loop and 2-loop level (see~(\ref{eq:tension},\ref{eq:i1})): 
\bea 
\ln \left[ \sigma a^2 \left(\beta\right) \right]_\mathrm{asym} 
^\mathrm{1-loop} &=& - \;
\frac{4 \pi^2}{\beta_0} \, [ \beta \, - \, \beta _\mathrm{ref}] 
\label{eq:as2} \\
&+& \ln \left[ \sigma a^2 \left(\beta_\mathrm{ref}\right) \right] 
\nonumber \\ 
\ln \left[ \sigma a^2 \left(\beta\right) \right]_\mathrm{asym} 
^\mathrm{2-loop} 
&=& - \;
\frac{4 \pi^2}{\beta_0} \, [ \beta \, - \, \beta _\mathrm{ref}] 
\label{eq:as3} \\ 
&+& \frac{2 \beta_1}{\beta_0^2} \, 
\ln  \frac{ \beta }{\beta _\mathrm{ref}}  \; + \; 
\ln \left[ \sigma a^2 \left(\beta_\mathrm{ref}\right) \right] 
 \; . 
\nonumber 
\ena 
The perturbative scaling functions are normalised to reproduce 
the measured data for $\beta = \beta_\mathrm{ref}$. In figure~\ref{fig:as1}, 
the highest considered value $\beta $ is chosen for $ \beta_\mathrm{ref}$. 
It is remarkable that both for SU(2) and SU(3), the 2-loop 
scaling function (\ref{eq:as3}) does not yield an improvement on 
the agreement of the data with the 1-loop formula (\ref{eq:as2}). 

\subsection{ Improved action }

Employing the  procedure discussed in subsection~\ref{sec:therm}, 
we generated well ``thermalized'' configurations (as well as the 
simulation parameters $w_{11}(\beta )$ and $w_{22}(\beta )$), see 
(\ref{eq:t3}), for a range of $\beta $ values which give reasonably sized 
lattice spacings for the present $16^4$ lattice. The simulation parameters 
as well as the calculated value of the lattice spacing $a$ in units of the 
string tension $\sigma $ are summarised in table~2 for the 
SU(2) gauge theory and in table~3 for the SU(3) case. 
The calculated scaling functions $\sigma a^2(\beta )$ are shown in 
figure~4 and figure~5, respectively. 
As with the Wilson action, a comparison with the asymptotic 
scaling functions (\ref{eq:as2},\ref{eq:as3}) is made. An 
satisfactory agreement with asymptotic scaling on coarse lattices  
with $\sigma a^2$ as large as $\sigma a^2 \approx 0.1$ is observed 
for both gauge groups. In both cases, the agreement with 
the 2-loop formula seems to be better than with the 1-loop result for 
$\sigma a^2 \le 0.06$.

\begin{center}
\begin{tabular}{c|cccl} 
$\beta $ & $ w_{11}(\beta )/2$ & $ w_{22}(\beta )/2$ & $N_\mathrm{conf}$ 
 & $\sigma a^2 $ \\ \hline 
         $1.250$ & $   0.62455(3)$ & $   0.13499(6) $ & $400$ &  $0.279(2)$   \\  
         $1.275$ & $   0.63416(3)$ & $   0.15944(6) $ & $400$ &  $0.215(5)$   \\  
         $1.300$ & $   0.64197(3)$ & $   0.17723(6) $ & $400$ &  $0.175(3)$   \\  
         $1.325$ & $   0.64885(3)$ & $   0.19211(6) $ & $400$ &  $0.1473(7)$  \\  
         $1.350$ & $   0.65526(3)$ & $   0.20584(6) $ & $400$ &  $0.1244(7)$  \\  
         $1.375$ & $   0.66123(3)$ & $   0.21818(6) $ & $400$ &  $0.1068(9)$  \\  
         $1.400$ & $   0.66690(3)$ & $   0.22986(6) $ & $400$ &  $0.0922(7)$  \\  
         $1.425$ & $   0.67231(3)$ & $   0.24121(6) $ & $400$ &  $0.0787(5)$  \\  
         $1.450$ & $   0.67744(3)$ & $   0.25169(6) $ & $400$ &  $0.0695(6)$  \\  
         $1.475$ & $   0.68229(3)$ & $   0.26143(6) $ & $400$ &  $0.0599(3)$  \\  
         $1.500$ & $   0.68697(2)$ & $   0.27076(6) $ & $400$ &  $0.0528(3)$  \\  
         $1.525$ & $   0.69141(2)$ & $   0.27964(6) $ & $800$ &  $0.0452(8)$  \\  
         $1.550$ & $   0.69572(2)$ & $   0.28809(6) $ & $800$ &  $0.0400(3)$  \\  
         $1.575$ & $   0.69988(2)$ & $   0.29638(6) $ & $800$ &  $0.0351(4)$  \\  
         $1.600$ & $   0.70388(2)$ & $   0.30427(6) $ & $800$ &  $0.0311(2)$
\end{tabular} 
\end{center}
%\medskip 
Table 2: Simulation parameters of the improved 
action (\ref{eq:t2}) and the calculated scaling function $\sigma a^2(\beta )$;
{\bf SU(2)} gauge theory, $16^4$ lattice, $N_\mathrm{conf}$ independent 
configurations. 

\centerline{  
\epsfxsize=8cm 
\epsffile{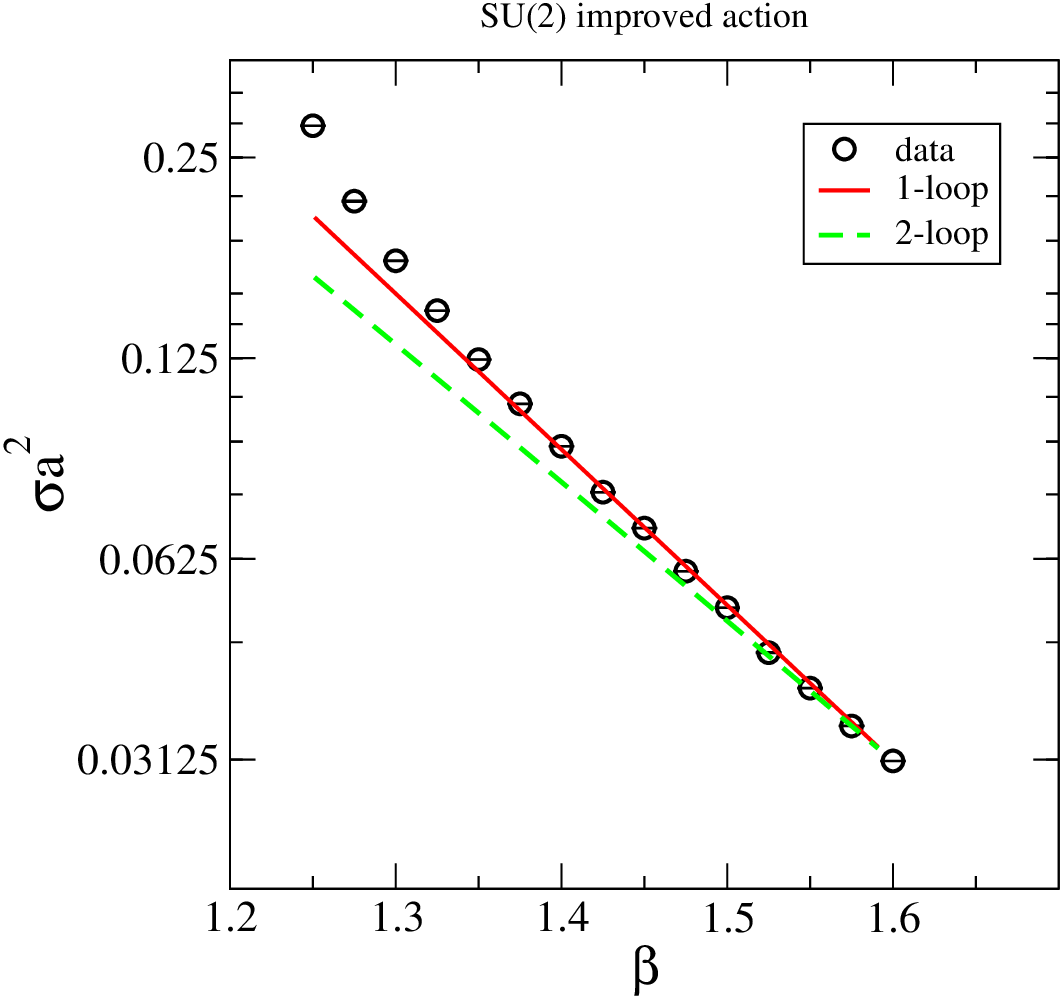} 
} 
\vskip 0.3cm
Figure 4: Scaling function $\sigma a^2(\beta)$ 
for a $16^4$ lattice using the present improved action for {\bf SU(2)}. 
Asymptotic scaling according to (\ref{eq:as2},\ref{eq:as3}) is shown 
as well. 

\begin{center}
\begin{tabular}{c|cccl} 
$\beta $ & $w_{11}(\beta )/3$ & $w_{22}(\beta )/3$ & $N_\mathrm{conf}$ 
 & $\sigma a^2 $ \\ \hline 
         $2.90$  &  $  0.58567(2) $ & $0.11424(3)$ & $800$ & $0.231(4) $ \\ 
         $3.00$  &  $  0.60135(2) $ & $0.14553(3)$ & $800$ & $0.151(2) $ \\ 
         $3.10$  &  $  0.61378(2) $ & $0.16726(3)$ & $800$ & $0.1122(7)$ \\ 
         $3.15$  &  $  0.61923(2) $ & $0.17667(3)$ & $800$ & $0.0985(5)$ \\ 
         $3.20$  &  $  0.62450(2) $ & $0.18560(3)$ & $800$ & $0.0851(4)$ \\ 
         $3.25$  &  $  0.62950(2) $ & $0.19403(3)$ & $800$ & $0.0765(4)$ \\ 
         $3.30$  &  $  0.63429(2) $ & $0.20215(3)$ & $800$ & $0.0666(3)$ \\ 
         $3.35$  &  $  0.63883(2) $ & $0.20959(3)$ & $800$ & $0.0589(3)$ \\ 
         $3.40$  &  $  0.64330(2) $ & $0.21711(3)$ & $800$ & $0.0532(2)$ \\ 
         $3.45$  &  $  0.64764(2) $ & $0.22432(3)$ & $800$ & $0.0482(2)$ \\ 
         $3.50$  &  $  0.65173(2) $ & $0.23108(3)$ & $800$ & $0.0424(2)$ 
\end{tabular} 
\end{center}
\medskip 
Table 3: Simulation parameters of the improved 
action (\ref{eq:t2}) and the calculated scaling function $\sigma a^2(\beta )$;
{\bf SU(3)} gauge theory, $16^4$ lattice, $N_\mathrm{conf}$ independent 
configurations. 

\vskip 0.8cm
\centerline{  
\epsfxsize=8cm 
\epsffile{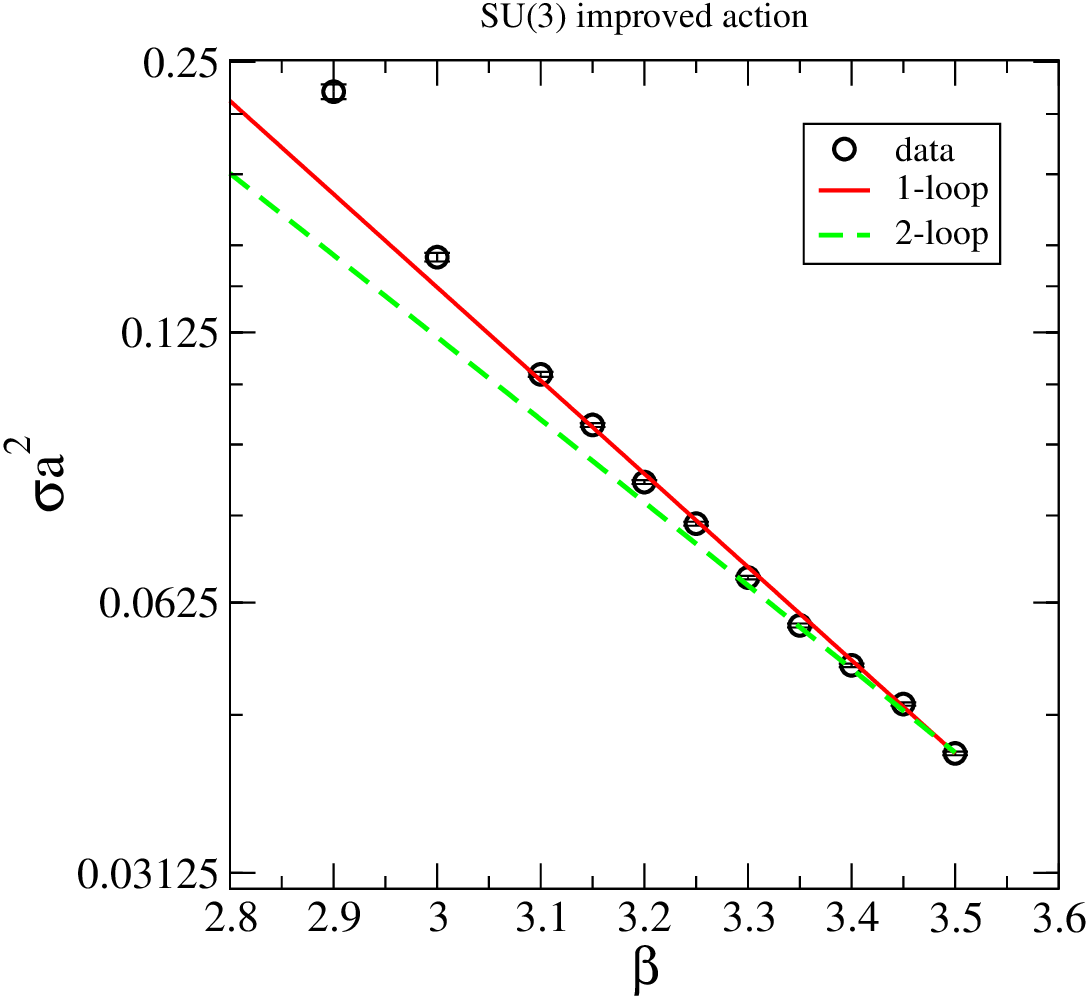} 
} 
\vskip 0.3cm
Figure 5: Scaling function $\sigma a^2(\beta)$ 
for a $16^4$ lattice using the present improved action for {\bf SU(3)}. 
Asymptotic scaling according to (\ref{eq:as2},\ref{eq:as3}) is shown 
as well. 

\subsection{ Comparison with other actions}

In this subsection, the more important case of a SU(3) gauge group 
is investigated. 
Two popular actions which do not involve tadpole improvement, but 
invoke a renormalisation group investigation, are the RG-Iwasaki 
action~\cite{Iwasaki:1985we,Iwasaki:1996sn} and the 
DBW2~\cite{deForcrand:1999bi}. These actions are of the type 
\bea 
S[U] &=& \beta \; \sum _{\mu < \nu, x} 
\Bigl[ c_0 \; \frac{1}{N_c}\tre \, W_{\mu \nu }^{1 \times 1} (x) 
\nonumber \\ 
&+& c_1 \; \frac{1}{N_c} \tre \, W_{\mu \nu }^{1 \times 2} (x) 
\Bigr] \; ,  
\label{eq:c1}
\ena 
and differ by the choice of $c_1$ (Note that $c_0 = 1 - 8 c_1$ 
for a proper definition of the bare gauge coupling):

\begin{center}
\begin{tabular}{ll}
$c_1 \; \approx \; -0.331 $ \hspace{2cm} & (RG-Iwasaki) \\
$c_1 \; \approx \; -1.4088 $ \hspace{2cm} & (DBW2). 
\end{tabular} 
\end{center} 

A detailed investigation of the scaling properties of these actions can 
be found in~\cite{Necco:2003vh}. We will here focus on their properties 
concerning {\it asymptotic scaling}. 

We will need the lattice spacing $a$ in units of the string tension. 
For the case of the RG-Iwasaki action and the DBW2 action, data for 
$a/r_0$ with the Sommer parameter $r_0$ are taken from the work by 
Necco~\cite{Necco:2003vh}. 
Using $r_0 \approx 0.5 \,$fm and $\sqrt{\sigma } \approx 440 \, $MeV,  
a factor 
$$ 
\sigma \; r_0^2 \; \approx \; 1.21 
$$
is used to convert $a^2/r_0^2$ to $a^2 \sigma $.

\vskip 0.3cm 
In order to study whether the present improved action (see equations 
(\ref{eq:t1}-\ref{eq:t3})) is superior to an action with standard 
tadpole improvement, we here also study the ``$2\times 2$'' action 
with tree-level coefficients and standard tadpole removal: 
\bea 
S[U](w_{11}) &=& \beta \; \sum _{\mu < \nu, x} 
\Bigl[ \frac{4}{3 \, w_{11} (\beta )} \; \tre \, W_{\mu \nu }^{1 \times 1} (x) 
\nonumber  \\
&-& \frac{1}{48 \, w_{11}^2 (\beta )} \; \tre \, 
W_{\mu \nu }^{2 \times 2} (x) \Bigr] \; , 
\label{eq:c2}
\ena 
where $w_{11}$ must be self-consistently calculated from 
\be 
w_{11} (\beta ) \; = \;  \frac{1}{Z} \int {\cal D} U_\mu \; 
\tre \, W_{\mu \nu }^{1 \times 1} (x) \; \exp \{ S[U](w_{11}) \} \; . 
\label{eq:c3} 
\en 
This action was used in~\cite{Beinlich:1995ik} to study 
thermodynamics. There it was observed that tadpole improvement 
largely reduces the cutoff effects which hamper the calculation of the 
pressure and the thermal energy density in the SU(3) high temperature phase. 
Our findings for $w_{11} (\beta ) $ and for the scaling function 
$\sigma a^2 (\beta) $ are summarised in table~\ref{tab:c1}. 

\setcounter{table}{3}

\begin{table}[h]
\begin{center}
\begin{tabular}{c|ccl} 
$\beta $ & $3 \, w_{11}(\beta )$ & $N_\mathrm{conf}$ 
 & $\sigma a^2 $ \\ \hline 
         $2.60$  & $0.57313(3) $  &  $600$ & $0.115(1)  $ \\ 
         $2.70$  & $0.58559(2) $  &  $600$ & $0.0875(6) $ \\ 
         $2.80$  & $0.59715(2) $  &  $600$ & $0.0683(4) $  \\ 
         $2.90$  & $0.60783(2) $  &  $600$ & $0.0520(3) $ \\ 
         $3.00$  & $0.61784(2) $  &  $600$ & $0.0431(3) $ \\ 
\end{tabular} 
\caption{ \label{tab:c1} Simulation parameter of the {\it standard 
tadpole improved action} (\ref{eq:c2}) and the calculated scaling 
function $\sigma a^2(\beta )$;
{\bf SU(3)} gauge theory, $16^4$ lattice, $N_\mathrm{conf}$ independent 
configurations. 
}
\end{center}
\end{table}

For a more quantitative  investigation of {\it asymptotic scaling}, 
the deviation from asymptotic scaling is measured by the ratio: 
\be 
R(\beta ) \; = \; \frac{ a^2 (\beta ) }{ a^2_\mathrm{asym}(\beta ) } \; ,
\label{eq:c4} 
\en 
where the lattice spacing squared $a^2$ is either provided in units 
of the string tension or in units of the Sommer parameter 
$r_0$ (as e.g.~done in~\cite{Necco:2003vh}). The function 
$a^2_\mathrm{asym}(\beta )$ is provided at 2-loop level by (\ref{eq:as3}). 
Because the definition of $a^2_\mathrm{asym}(\beta )$ involves an 
arbitrary normalisation, the absolute value of $R$ in (\ref{eq:c4}) 
is meaningless. The data are normalised such that $R=1$ is attained 
for the smallest lattice spacing considered. 
Asymptotic scaling will from  (\ref{eq:c4}) be signalled by 
the function $R(\beta )$ becoming flat for sufficiently large 
values of $\beta $. Since the absolute size of the bare gauge coupling $g$
(and therefore of $\beta =6/g^2$) depends on the details of the regularisation 
scheme and the action, we will study $R$ as a function of 
the lattice spacing squared in physical units.

\vskip 0.3cm 
\begin{figure}[t]
\centerline{  
\epsfxsize=8cm 
\epsffile{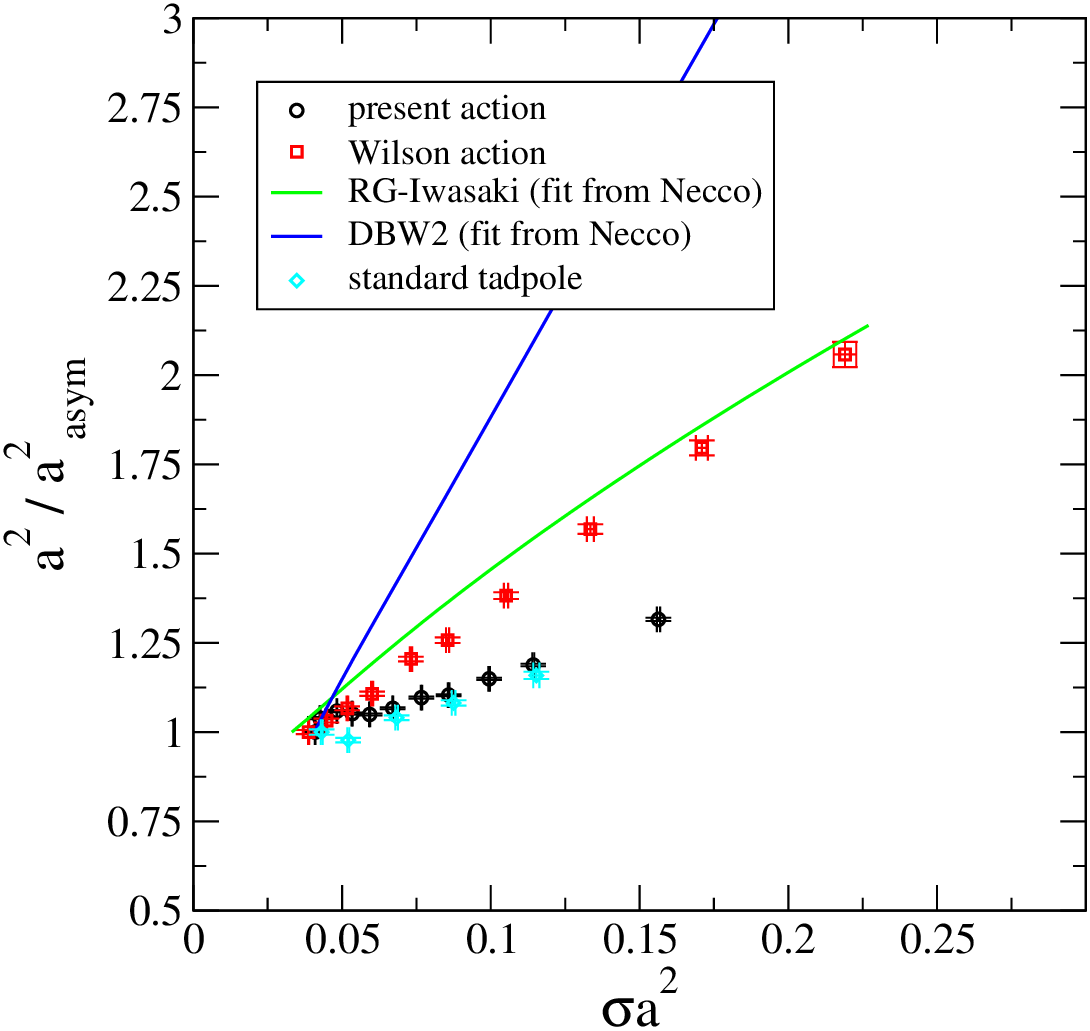} 
} 
\vskip 0.3cm
\caption{ \label{fig:asym} Approaching asymptotic scaling using several 
actions. 
}
\end{figure}
The results are shown in figure~\ref{fig:asym} for the actions under 
investigations. It turns out that any sort of tadpole improvement 
largely improves scaling along the lines of the asymptotic formula. 
The Wilson action, but also the DBW2 and Iwasaki actions, show large 
deviations from asymptotic scaling. In contrast, both, the standard 
and our new tadpole improved action, seem to perform equally well. 
A possible explanation could be that both actions remove the 
$O_6$ irrelevant terms such that asymptotic scaling 
sets in for rather coarse lattice spacings. The next section will, however, 
reveal that this interpretation is only justified for the new 
action proposed in this paper.

\section{ Rotational symmetry breaking \label{sec:rot} }

The irrelevant terms of $O_6$ in (\ref{eq:ai5}) are built up from 
expressions which explicitly violate rotational symmetry. 
Hence, the absence of the  $O_6$ terms can be checked 
by calculating the amount (for a proper definition see below)
of rotational symmetry breaking as a function 
of the lattice spacing $a$. 

\vskip 0.3cm 
The present improvement scheme belongs to the class of tree level 
improvements: it relies on the expansion (\ref{eq:ai6})
of the action in powers of the lattice spacing $a$. 
A cancellation of the $O_6$ terms can be hampered by terms which 
depend logarithmically on $a$. The cancellation can still be made 
complete if loop corrections are considered as well as tadpole 
improvement. At the present stage, there are two crucial questions: 
Are loop corrections still large for the present range of lattice 
spacings? Is the new improvement scheme superior to the standard 
approach so that the additional complexity of the new scheme 
is justified?

\vskip 0.3cm 
For an answer to these question, we need to quantify the amount of 
rotational symmetry breaking. For this purpose, we invoke 
the method introduced by the QCD Taro collaboration 
in~\cite{deForcrand:1999bi}. 
Let $V_\mathrm{on}(r)$ denote the ``on-axis'' static quark potential 
obtained from quarks positioned along the main crystallographic direction 
previously called the $(100)$ direction. Data for which $r$ is not an 
integer multiple of $a$ are made available by means of the fit 
(\ref{eq:p2}). Let furthermore call $V(r)$ all data 
arising from quarks positioned along the $(110)$ and $(111)$ directions. 
These data are called the ``off-axis'' data. 
$\delta V(r)$ denotes their statistical errors. 
With these definitions, the measure of rotational symmetry breaking
is given by: 
\be 
\delta _v^2 \; = \; \sum _{\mathrm{off}} \frac{ 
[ V(r) \; - \; V_\mathrm{on}(r) ]^2 }{ V(r)^2 \; \delta V ^2 (r) } 
\; \big/ \; \bigl( \sum _{\mathrm{off}} 
\frac{1}{  \delta V ^2 (r) } \bigr) \; ,  
\label{eq:del}
\en
where the sum extends over all ``off-axis'' data. 

\setcounter{figure}{5}
\begin{figure}[t]
\centerline{  
%\epsfxsize=8cm 
%\epsffile{rot_su2.eps} \hspace{.5cm}
\epsfxsize=8cm 
\epsffile{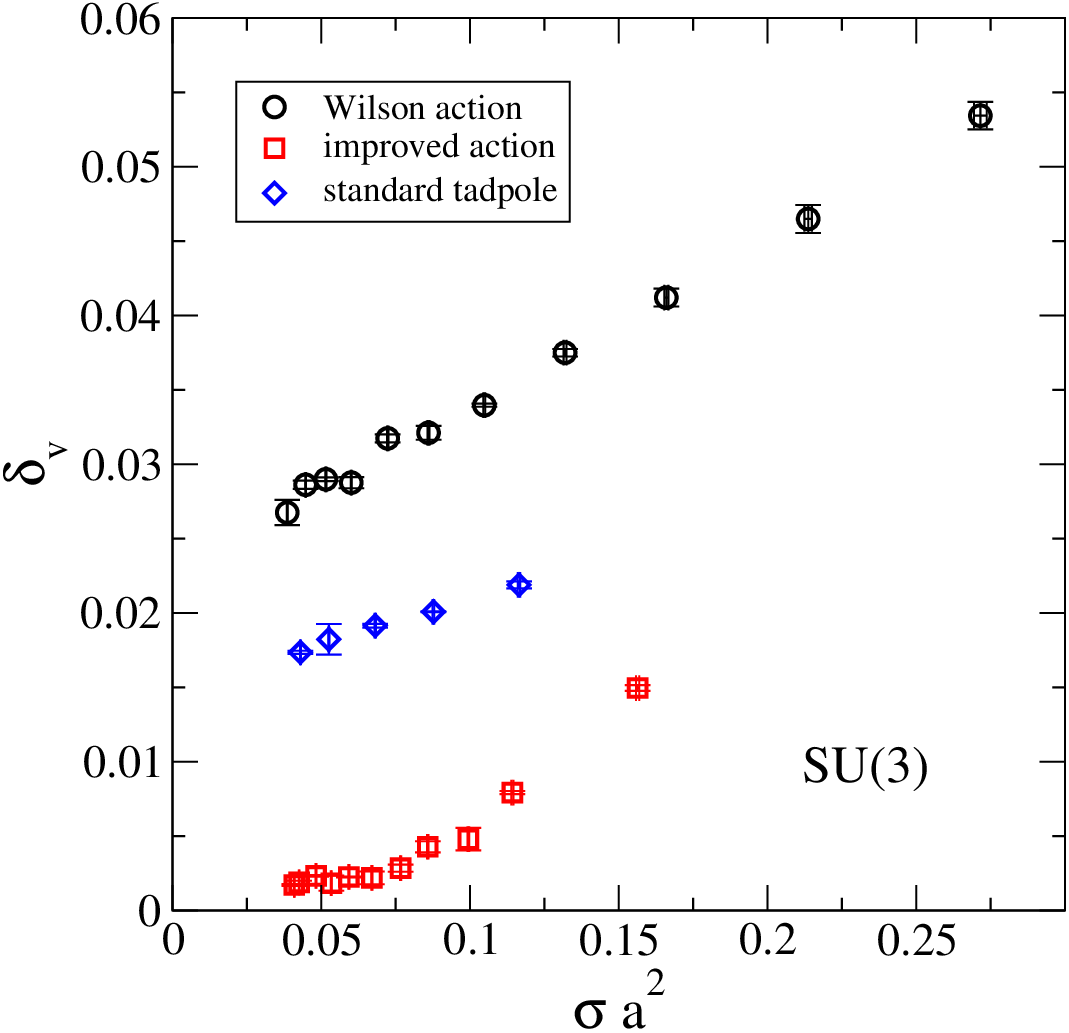} 
} 
\vskip 0.3cm
\caption{ \label{fig:rot} The measure $\delta _v$ of rotational 
symmetry breaking as a function of the scaling function $\sigma a^2$ 
for SU(3) gauge theory with different actions. 
}
\end{figure}
Figure~\ref{fig:rot} shows $\delta _v$ as a function of the lattice spacing 
squared. For the case of the Wilson action, we observe a linear scaling 
of $\delta _v$ with $a^2$: 
$$ 
\delta _v \; \propto \; \sigma a^2 \; , \hbo \hbox{(Wilson action)} \; . 
$$
In the case of the action employing 
standard tadpole improvement, the symmetry breaking effects are 
significantly reduced, but, still, $\delta _v$ rises linearly with $a^2$. 
Using the new action proposed in this paper, we find an additional 
drastic reduction of rotational symmetry breaking effects. 
Moreover, it seems that this time the functional dependence 
of $\delta _v$ on $a^2$ seems to change qualitatively:  
The data now indicate that $\delta _v $ is of higher order in $a^2$: 
$$ 
\delta _v \; \propto \; \sigma^2 a^4 \; \; , \hbo \hbox{(SU(3) imroved 
action)} \; . 
$$
However, further numerical investigations are necessary to support this 
claim. If it is supported by numerical simulations, 
it would imply that only the present action can completely  
remove the irrelevant  $O_6$ contributions. It is already clear that 
the standard tadpole improved action certainly fails this task. 

\vskip 0.3cm 
Note also that the data suggest that the Wilson and standard tadpole 
data approach a small but finite value at $\sigma a^2=0$. 
The size of this value clearly depends on the amount of tadpole 
contributions. Our preliminary interpretation of this finding is as follows: 
contributions of tadpole loops solely arise in lattice regularisation and, 
therefore, add substantially to the amount of rotational symmetry breaking 
present in the static potential. In addition, tadpole loops 
are generically UV divergent. This might lead to a small, but finite 
value of $\delta _v$ even for very small values of the lattice spacing. 
Further investigations are clearly needed to settle this question.

\section{Conclusions} 

The properties of improved actions with respect to asymptotic scaling 
has been thoroughly investigated in this paper. A focal point of the 
present study is tadpole improved tree level actions. 
A new scheme for tadpole improvement has been proposed and it has been 
contrasted to the heuristic tadpole approach, which is 
standard in the literature. It has been shown that the standard tadpole 
scheme is a mean field approximation to the scheme proposed here. 

\vskip 0.3cm 
The numerical results for the scaling function $\sigma a^2(\beta )$ 
reveal that both types of tadpole improved actions yield results of 
equal quality as far as asymptotic scaling is concerned. 
By contrast, loop improved actions (which do not make use of tadpole 
improvement) produce much bigger deviations from asymptotic scaling. 

\vskip 0.3cm 
The amount $\delta _v$  of rotational symmetry breaking (see 
(\ref{eq:del}))in the static quark potential was used to compare the quality 
of both tadpole improvement schemes. Although the function $\delta _v (a) $ 
is much smaller for the standard tadpole action than for the Wilson action, 
the functional dependence on the lattice spacing $a$ is the same in 
both cases. In contrast, we have seen first numerical evidence that 
$\delta _v$ is of higher order in $a^2$ if the new tadpole improved 
action is used. We here argue that the generic tadpole scheme fails 
to eliminate the leading order irrelevant terms of the action.
The data indicate that the new action cancels these terms from 
the action for the range of lattice spacings considered 
without taking into account loop corrections. 

\vskip 0.3cm 
Our approach to tadpole improvement can in principle be 
extended to kill off  the next to leading order irrelevant terms 
as well. The question whether loop corrections must be considered then 
is left to future work.

\bigskip 
{\bf Acknowledgements: } We thank Tom Heinzl and Martin Lavelle for 
helpful comments on the manuscript. 
Numerical simulations have been carried out using the HPC facility 
{\tt PlymGrid} in Plymouth. We thank the staff for support.

\end{document}